\documentclass[journal]{IEEEtran}
\usepackage{cite}%

\ifCLASSINFOpdf
\else
\fi
\usepackage{amssymb,amsmath}
\usepackage{amsfonts,mathrsfs}

\usepackage{bm}
\newcommand{\bs}[1]{\bm{\mathrm{#1}}}

\usepackage[g]{esvect}

\newcommand{\vect}[1]{\overline{#1}}
\newcommand{\nhat}{\ensuremath{\hat{n}}}

\newcommand{\vecprime}[1]{\vect{#1}^{\,\prime}}

\renewcommand{\r}{\left(\vect{r}\right)}

\newcommand{\rrpki}[1][]{\left(k_i, \vect{r}, \vecprime{r}\right)}

\newcommand{\matr}[1]{\bs{#1}}

\newcommand{\junk}[1] {}

\def\XXint#1#2#3{{\setbox0=\hbox{$#1{#2#3}{\int}$}
\vcenter{\hbox{$#2#3$}}\kern-.5\wd0}}

\newcommand*\widebar[1]{%
  \hbox{%
    \vbox{%
      \hrule height 0.5pt 
      \kern0.3ex
      \hbox{%
        \kern-0.05em
        \ensuremath{#1}%
        \kern-0.05em
      }%
    }%
  }%
} 

\renewcommand{\epsilon}{\varepsilon}

\newcommand{\surf}{\mathcal{S}}

\newcommand{\Lmat}[1][]{{\matr{L}_{#1}}}
\newcommand{\LmatA}[1][]{{\matr{L}^{(A)}_{#1}}}
\newcommand{\LmatPhi}[1][]{{\matr{L}^{(\phi)}_{#1}}}
\newcommand{\Kmat}[1][]{{\matr{K}_{#1}}}

\newcommand{\Pxout}{\ensuremath{\matr{I}_{\times}}} 

\newcommand{\figref}[1]{Fig.~\ref{#1}}
\newcommand{\secref}[1]{Section~\ref{#1}}
\newcommand{\tblref}[1]{Table~\ref{#1}}

\usepackage{tcolorbox}

\usepackage{textpos}

\newcommand{\mySubtitle}[1]%
{%
	\begin{textblock}{14.0}(0.7, 2.9)
		\textbf{#1}%
	\end{textblock}%
}%

\newcommand{\redcol}{black}

\newcommand{\red}[1]{\textcolor{\redcol}{#1}}
\newcommand{\green}[1]{\textcolor{green!0!black}{#1}}

\newcommand{\Ecolor}[1]{{#1}}
\newcommand{\Hcolor}[1]{{#1}}
\newcommand{\Jcolor}[1]{{#1}}

\newcommand{\Er}[1][]{\Ecolor{\vect{E}_{#1}\r}}

\newcommand{\Etr}[1][]{\Ecolor{\nhat_{#1} \times \vect{E}_{#1}\r}}

\newcommand{\Emat}[1][]{\Ecolor{\matr{E}_{#1}}}

\newcommand{\Hr}[1][]{\Hcolor{\vect{H}_{#1}\r}}

\newcommand{\Htr}[1][]{\Hcolor{\nhat_{#1} \times \vect{H}_{#1}\r}}

\newcommand{\Hmat}[1][]{\Hcolor{\matr{H}_{#1}}}

\newcommand{\Jmat}[1][]{\ensuremath{\Jcolor{\matr{J}_{#1}}}}

\newcommand{\Grrpki}[1][]{\ensuremath{\green{G\rrpki}}}

\newcommand{\rhomat}[1][]{\red{\matr{\rho}_{#1}}}

\makeatletter
\newlength\numerator@height
\newlength\frac@height
\newsavebox\numerator@box
\newsavebox\frac@box

\newcommand\dfracparens[3]{%
	\sbox{\numerator@box}{\ensuremath{#1}}%
	\sbox{\frac@box}{\ensuremath{\dfrac{#1}{#2}}}%
	\settoheight{\frac@height}{\usebox{\frac@box}}%
	\settoheight{\numerator@height}{\usebox{\numerator@box}}%
	\addtolength{\frac@height}{-\numerator@height}%
	\usebox{\frac@box}%
	\raisebox{\frac@height}{%
		\( \left( {#3} \right)
		\)}%
}
\makeatother

\usepackage{hhline}
\usepackage{lipsum} 
\makeatletter
\newcommand{\removelatexerror}{\let\@latex@error\@gobble}
\makeatother

\usepackage{algorithm}
\usepackage{algpseudocode}
\newsavebox{\ieeealgbox}

\ifCLASSOPTIONcompsoc
  \usepackage[caption=false,font=normalsize,labelfont=sf,textfont=sf]{subfig}
\else
  \usepackage[caption=false,font=footnotesize]{subfig}
\fi
\begin{document}
%
\title{A Parallel Boundary Element Method for the Electromagnetic Analysis of Large Structures With Lossy Conductors}
%
%
%

\author{Damian~Marek,~\IEEEmembership{Graduate~Student~Member,~IEEE,}
        Shashwat~Sharma,~\IEEEmembership{Graduate~Student~Member,~IEEE,}
        and~Piero~Triverio,~\IEEEmembership{Senior~Member,~IEEE}
\thanks{D. Marek and S. Sharma are with the Edward S. Rogers Sr. Department of Electrical \& Computer Engineering, University of Toronto, Toronto, ON, M5S 3G4 Canada, P. Triverio is with the Edward S. Rogers Sr. Department of Electrical \& Computer Engineering and with the Institute of Biomedical Engineering, University of Toronto, e-mails: damian.marek@mail.utoronto.ca, shash.sharma@mail.utoronto.ca, piero.triverio@utoronto.ca.}
\thanks{Manuscript received Month X, 2021; revised Month X, 2021.}}

%
%

\markboth{IEEE Transactions on Antennas and Propagation,~Vol.~x, No.~y, Month~2021}%
{Shell \MakeLowercase{\textit{et al.}}: Bare Demo of IEEEtran.cls for IEEE Journals}
%



\maketitle

\begin{abstract}
In this paper, we propose an efficient parallelization strategy for boundary element method (BEM) solvers that perform the electromagnetic analysis of structures with lossy conductors. The proposed solver is accelerated with the adaptive integral method, can model both homogeneous and multilayered background media, and supports excitation via lumped ports or an incident field. Unlike existing parallel BEM solvers, we use a formulation that rigorously models the skin effect, which results in two coupled computational workloads. The external-problem workload models electromagnetic coupling between conductive objects, while the internal-problem workload describes field distributions within them. We propose a parallelization strategy that distributes these two workloads evenly over thousands of processing cores. The external-problem workload is balanced in the same manner as existing parallel solvers that employ approximate models for conductive objects. However, we assert that the internal-problem workload should be balanced by algorithms from scheduling theory. The parallel scalability of the proposed solver is tested on three different structures found in both integrated circuits and metasurfaces. The proposed parallelization strategy runs efficiently on distributed-memory computers with thousands of CPU cores and outperforms competing strategies derived from existing methods.
\end{abstract}


%
\IEEEpeerreviewmaketitle

\section{Introduction}\label{sec:intro}
\IEEEPARstart{A}{ccurately} modeling the electromagnetic properties of lossy conductors is needed to design advanced devices in electrical engineering. For example, in metamaterials, the subwavelength feature sizes and sharp corners of metallic scatterers require accurate modeling so that ohmic losses can be successfully predicted~\cite{Guney2009}. These issues are also present when modeling signal propagation and coupling in the complex layouts of high-speed electrical interconnects. In fact, the electromagnetic response of electrical interconnects is needed over a wide frequency band where large variations in skin depth occur~\cite{Paul2007, Qian2007}.

In addition, metamaterials and electrical interconnects are both multiscale structures that result in large and computationally expensive problems that need to be solved with efficient numerical methods. A strong candidate for these problems is the boundary element method (BEM). The BEM's advantage is that it only requires a 2D mesh of the surface of each object, which normally results in fewer degrees of freedom than volumetric approaches, such as the finite element method~\cite{Jin2014} or volumetric integral equations~\cite{Omar2013}. Furthermore, the BEM tends to be more effective when lossy conductors are considered, because volumetric methods require an extremely fine mesh inside conductive objects to accurately capture rapid electromagnetic field variations due to the skin effect.

Unfortunately, the BEM results in a dense system of equations whose direct solution is only tractable for small problems. Most often, the BEM is partnered with an acceleration technique, for instance, the multilevel fast-multipole algorithm (MLFMA)~\cite{Song1995}, the adaptive integral method (AIM)~\cite{Bleszynski1996}, or hierarchical matrices~\cite{Bebendorf2008}. Then, an iterative method is used to solve the system of equations, e.g., the generalized minimal residual (GMRES) method~\cite{Saad1986}. The AIM is more commonly used in situations involving layered media, because modifying the MLFMA for layered media is not easy~\cite{AIM_MGF_2D, okh_AIM_MGF_3D, convcorr_AIM}. Despite the algorithmic enhancements available for the BEM, many practical problems are still too computationally expensive to be solved on single workstations. Thus, parallelization methods are needed to formulate efficient electromagnetic solvers for distributed-memory supercomputers. In a distributed-memory setting, information must be communicated between different compute nodes, which can become a bottleneck. Therefore, an important goal of distributed-memory algorithms is to minimize the amount of communication required.

In the literature, there are numerous parallel MLFMA-based BEM solvers for modeling extremely large objects~\cite{Velamparambil2003, Ergul2008, Pan2012, Taboada2013}, but they are all restricted to perfect electric conductors~(PECs). Parallelization methods have also been investigated for other acceleration techniques, e.g., the AIM \cite{Phillips1997, Wei2011} and the adaptive cross approximation (ACA) algorithm \cite{Schroder2014,He2015}, these works were also focused on modeling PEC objects. For layered media, \cite{Liu2019} and \cite{Marek2020b} have presented parallel AIM solvers that model lossy conductors through the surface impedance boundary condition (SIBC)~\cite{Yuferev2009}, which is accurate only when skin effect is sufficiently developed. Parallel BEM solvers that accurately model penetrable objects are rare, yet there are examples of parallel MLFMA and ACA solvers that target dielectric objects~\cite{Fostier2008,Ergul2011,Ergul2013,Gibson2020}. However, these solvers make use of the Poggio-Miller-Chang-Harrington-Wu- Tsai (PMCHWT)~\cite{Poggio1973, Chang1977, Wu1977} formulation, or related formulations, which are challenging to apply to conductive objects, because of the large contrast in electrical properties with the surrounding medium~\cite{Qian2007}. To the best of our knowledge, there are no parallelization strategies in the literature for BEM solvers that accurately model electromagnetic phenomena inside of lossy conductors, such as skin and proximity effect.

The main goal of this paper is to address this gap by introducing a parallelization strategy for the single layer impedance matrix (SLIM) formulation \cite{Sharma2020_slim}, which has been shown to facilitate the modeling of conductors embedded in a layered medium~\cite{Sharma2020_slim}. In the SLIM formulation, the electromagnetic response of metallic objects embedded in a background medium is modeled by two coupled problems. The external problem describes electromagnetic interactions between conductive objects through the surrounding medium. Instead, the internal problem describes the electromagnetic field distribution inside conductive and dielectric objects, capturing phenomena like skin effect and ohmic losses. The external and internal problems are each associated with a computational workload, which for brevity will be referred to as the external workload and the internal workload. Devising an efficient parallelization strategy for these coupled workloads is non-trivial because of the following reasons:
\begin{enumerate}
	\item The external workload is a single large computational task, which should ideally be performed by all processors.
	\item On the other hand, the internal workload consists of several independent tasks that are associated to individual conductive objects. Each internal workload should ideally use a number of processors adequate to the electrical size and complexity of the object. 
	\item Assigning processors to the internal workloads must be done carefully to ensure that the computations associated to each object can proceed independently, without having some processors sit idle while waiting for other required processors to finish previous tasks. In addition, a balancing strategy is needed to assign processors an equal amount of computational work, which in itself is a non-trivial problem requiring algorithms from scheduling theory. In fact, distributing a set of discrete parallel computational tasks over an arbitrary number of identical processing units is a variant of the multiprocessor scheduling problem~\cite{Drozdowski2009}. This class of problems is known to be strongly nondeterministic polynomial-time (NP)-hard~\cite{Du1989}. 
	\item Processors must be carefully assigned to these two workloads since the internal and external problems are coupled and share the same set of unknowns. Assigning the internal and external workloads completely independently could result in excessive communication, which would diminish the solver's parallel efficiency.
\end{enumerate}

In this paper, we address these challenges by proposing an efficient parallelization strategy for BEM algorithms that solve Maxwell's equations with accurate modeling of lossy conductors. A greedy list scheduling algorithm is proposed to distribute the internal workloads in an efficient manner, assigning an adequate number of processors to each object and minimizing idle time~\cite{Graham1966, Graham1969, Belkhale1990}. The external workload is instead parallelized using a graph partitioning approach~\cite{Marek2020a,Marek2020b}, which provides flexibility to map processors to the external and internal workloads in a coherent fashion that minimizes communication. Although the proposed parallel solver is developed for the SLIM formulation, the underlying algorithm for distributing internal workloads could easily be paired with other BEM techniques~\cite{Qian2007}.

The paper is organized as follows. In \secref{sec:slim}, the BEM formulation that we adopt is briefly summarized. Next, in \secref{sec:prop-para}, we propose an efficient parallelization strategy, which uses techniques from scheduling theory, to efficiently solve the system of equations in \secref{sec:slim}. Finally, in \secref{sec:val-scala} a parallel solver based on the proposed strategy is validated and its parallel scalability characterized.
\section{SLIM Formulation}\label{sec:slim}%
We study the problem of modeling a conductive object embedded in a homogeneous or layered medium and excited by an incident field or by lumped ports~\cite{Wang2004}. For now a single object is considered. Later, the formulation will be extended to multiple objects. Formally, we consider a background layered medium defined as $\mathcal{V}_0 = \bigcup_{l=1}^{N_l}\mathcal{V}_l$, where the $l^{\mathrm{th}}$ layer has volume $\mathcal{V}_l$, electric permittivity $\epsilon_l$ and magnetic permeability $\mu_l$. Embedded in this background medium is a homogeneous object with permittivity $\epsilon$, permeability $\mu$, and electric conductivity $\sigma$. This object has a volume $\mathcal{V}$ bounded by a surface $\surf$ with outward unit normal vector $\nhat$. In addition, an excitation field $\Er[\mathrm{inc}]$ with $\vect{r} \in \mathcal{V}_0$ interacts with the conductor to produce a field distribution $\left[\Er, \Hr \right]$.

In the BEM, Maxwell's equations are expressed in terms of surface integral equations that involve the Green's function of the background medium~\cite{Gibson2014}. Several BEM formulations invoke the equivalence principle~\cite{Harrington2001} along with a known Green's function~\cite{Gibson2014} to create an equivalent problem where objects are replaced by the surrounding material. In this section, the SLIM formulation is summarized. For details the reader is referred to~\cite{Sharma2020_slim}. In the SLIM formulation, the conductive object is replaced by the material of the surrounding layer and equivalent electric surface currents $\vect{J}_{\Delta}\r$ are impressed that reproduce $\Er$ and $\Hr$ outside $\surf$, while maintaining continuity of $\nhat \times \Er$ on $\surf$. Inside $\surf$ the magnetic field is changed and represented by a new field $\Hr[\mathrm{eq}]$, which is related to $\vect{J}_{\Delta}\r$ as
\begin{align}
  \vect{J}_{\Delta}\r &= \Htr - \nhat \times \Hr[\mathrm{eq}].\label{eq:Jdelta}
\end{align}
Next, $\surf$ is meshed with triangular elements, and Rao–Wilton–Glisson (RWG) basis functions $\vect{f}_n\r$~\cite{Rao1982} are used to expand $ \vect{J}_{\Delta}\r$, $\nhat \times \Er$, $\Htr$, and $\nhat \times \Hr[\mathrm{eq}]$. Their associated unknowns are collected into vectors~$\Jmat[\Delta]$, $\Emat$, $\Hmat$, and $\Hmat[\mathrm{eq}]$, respectively. Equation \eqref{eq:Jdelta} can now be written in discretized form as
\begin{align}
  \Jmat[\Delta] = \Hmat - \Hmat[\mathrm{eq}].\label{eq:Jdeltadis}
\end{align}%

\subsection{External Problem}\label{sec:slim:external}
The equivalent problem consists of a current density distribution radiating in the background medium. Thus, $\Er[\mathrm{inc}]$ can be related to $\vect{J}_{\Delta}\r$ and $\nhat \times \Er$ on $\surf$ through the augmented electric field integral equation (AEFIE)~\cite{Qian2009}. In the AEFIE, an equivalent charge density $\rho_\Delta\r$ is introduced, expanded with pulse basis functions $h_n\r$, and then related to $\vect{J}_{\Delta}\r$ through the discretized continuity equation.

As a result, the AEFIE system of equations~\cite{Qian2009} is written as
\begin{align}
\begin{bmatrix}
jk_0 \LmatA[\mathrm{m}] &
\matr{D}^T\LmatPhi[\mathrm{m}]\matr{B} & \Pxout \\
\matr{F}\matr{D} & jk_0\matr{I} & 0
\end{bmatrix}
\begin{bmatrix}
\Jmat[\Delta] \\ c_0\rhomat[\Delta] \\ \Emat/\eta_0
\end{bmatrix} =
\begin{bmatrix}
\Emat[\mathrm{inc}]/\eta_0 \\ \matr{0}
\end{bmatrix},\label{eq:AEFIEsys}
\end{align}
where the first row is the electric field integral equation (EFIE) augmented with the term related to charge density, and the second row is the discretized continuity equation. In~\eqref{eq:AEFIEsys}, $k_0$, $c_0$, and $\eta_0$ are free space values of the wave number, speed of light, and wave impedance, respectively. Sparse matrices $\matr{F}$, $\matr{B}$ and $\matr{D}$ are defined in~\cite{Qian2009}, $\matr{I}$ is the identity matrix, and $\rhomat[\Delta]$ is the vector of unknowns associated with $\rho_\Delta\r$. Sparse matrix $\Pxout$ is an identity operator formed by testing $\vect{f}_n\r$ with rotated RWG functions $\nhat \times \vect{f}_m\r$, and the incident field is represented by $\Emat[\mathrm{inc}]$. Finally, $\LmatA[\mathrm{m}]$ and $\LmatPhi[\mathrm{m}]$ are dense matrices whose entries are computed through numeric integration involving testing functions, basis functions, and the multilayered Green's function~\cite{Michalski1990,Sharma2020_slim}.   

The system in~\eqref{eq:AEFIEsys} is underdetermined because of the extra unknown $\Emat$. If the object is a PEC,~\eqref{eq:AEFIEsys} can be solved by setting both $\Emat$ and $\Hmat[\mathrm{eq}]$ to $0$. If the object is lossy and the frequency is sufficiently high that the skin effect is well-developed, the SIBC~\cite{Yuferev2009} can be used to enforce a local approximate relationship between $\Emat$ and $\Jmat[\Delta]$ while setting $\Hmat[\mathrm{eq}]$ to 0, thus eliminating $\Emat$. However, if the skin effect is not well-developed or if regions of $\surf$ have high curvature, the SIBC may be inaccurate~\cite{Yuferev2009}. In these cases, accurately modeling the conductive object requires fully accounting for all electromagnetic interactions through the internal medium. To accomplish this, the SLIM formulation~\cite{Sharma2020_slim} introduces an internal problem, which relates $\Emat$ to $\Jmat[\Delta]$ through additional integral equations.

\subsection{Internal Problem}\label{sec:slim:internal}
The tangential magnetic field just inside $\surf$ can be related to $\Etr$ through rearranging the discretized magnetic field integral equation (MFIE)~\cite{Gibson2014,Sharma2020_slim}
\begin{align}
  \Emat = \underbrace{\dfrac{-1}{j\omega\epsilon} \left(\Lmat\right)^{-1} \left(\Kmat + \dfrac{1}{2}\Pxout\right)}_{\triangleq\matr{Z}}\Hmat,\label{eq:MFIEinZin}
\end{align}
where $\omega$ is the angular frequency, $\Lmat$ is the discretized single-layer potential operator, and $\Kmat$ is the discretized double-layer potential operator~\cite{Gibson2014}. The entries of $\Lmat$ and $\Kmat$ are computed with the homogeneous Green's function using the material properties of the conductive object, so specialized integration routines are needed when conductivity is high~\cite{Qian2007}. Using~\eqref{eq:MFIEinZin}, $\Emat$ can be expressed in terms of $\Hmat$ by an impedance matrix $\matr{Z}$.
	
Similarly, in the equivalent problem, the discretized EFIE can be invoked to relate $\Etr$ to $\nhat \times \Hr[\mathrm{eq}]$~\cite{Gibson2014,Sharma2020_slim}:  
\begin{align}
  \Hmat[\mathrm{eq}] = \underbrace{\dfrac{1}{j\omega\mu_{l}} \left(\matr{L}_{l}\right)^{-1} \left(\matr{K}_{l} + \dfrac{1}{2}\matr{I}_{\mathrm{x}}\right)}_{\triangleq\matr{Y}_{\mathrm{eq}}} \Emat,\label{eq:EFIEeqYeq}
\end{align}
where quantities with subscript $l$ depend on the material properties of the $l$-th layer~\cite{Sharma2020_slim}.

Combining~\eqref{eq:MFIEinZin} and~\eqref{eq:EFIEeqYeq}, we can relate $\Hmat[\mathrm{eq}]$ to $\Hmat$ as
\begin{align}
  \Hmat[\mathrm{eq}] = \matr{Y}_{\mathrm{eq}} \matr{Z} \Hmat.\label{eq:EFIEeqYZ}
\end{align}

\subsection{Final System Matrix}\label{sec:slim:system}
The final system of equations for the SLIM formulation~\cite{Sharma2020_slim} is arrived at by first substituting~\eqref{eq:MFIEinZin} into~\eqref{eq:AEFIEsys}. Next, equation~\eqref{eq:EFIEeqYZ} is substituted into~\eqref{eq:Jdeltadis}, and the result is combined with~\eqref{eq:AEFIEsys} to give
\begin{align}
  \begin{bmatrix}
    jk_0 \LmatA[\mathrm{m}] + \matr{C} &
    \matr{D}^T\LmatPhi[\mathrm{m}]\matr{B} \\
    \matr{F}\matr{D}\left(\matr{I} - \matr{Y}_{\mathrm{eq}}\matr{Z}\right) & jk_0\matr{I}
  \end{bmatrix}
  \begin{bmatrix}
    \Hmat \\ c_0\rhomat[\Delta]
  \end{bmatrix} =
  \begin{bmatrix}
    \Emat[\mathrm{inc}]/\eta_0 \\ \matr{0}
  \end{bmatrix},\label{eq:AEFIEoutJ}
\end{align}
where
\begin{align}
  \matr{C} = \left(-jk_0 \LmatA[\mathrm{m}] \matr{Y}_{\mathrm{eq}} + \eta_0^{-1}\Pxout\right)\matr{Z}.\label{eq:Cdef}
\end{align}

The SLIM formulation can be extended to multiple objects by defining $\surf = \bigcup_{i=1}^{N_{\mathrm{obj}}}\surf_i$, where $\surf_i$ is the surface enclosing the $i$-th object and $N_{\mathrm{obj}}$ is the total number of objects. In this case, the procedure in~\secref{sec:slim:internal} is repeated for each $\surf_i$, and the $i$-th object is associated with matrices $\matr{Z}_i$ and $\matr{Y}_{i,\mathrm{eq}}$~\cite{Sharma2020_slim}. Matrices $\matr{Z}$ and $\matr{Y}_{\mathrm{eq}}$ in~\eqref{eq:AEFIEoutJ} are then defined by a block diagonal concatenation of all $\matr{Z}_i$ and $\matr{Y}_{i,\mathrm{eq}}$ matrices~\cite{Sharma2020_slim},
\begin{align}
	\matr{Z} &= \mathrm{diag}\begin{bmatrix}\matr{Z}_1&\matr{Z}_2&\cdots &\matr{Z}_{N_\mathrm{obj}}\end{bmatrix},\\
	\matr{Y}_\mathrm{eq} &= \mathrm{diag}\begin{bmatrix}\matr{Y}_{1,\mathrm{eq}}&\matr{Y}_{2,\mathrm{eq}}&\cdots &\matr{Y}_{N_\mathrm{obj},\mathrm{eq}}\end{bmatrix}.
\end{align}

The majority of matrices in~\eqref{eq:AEFIEoutJ} are sparse with the exception of $\LmatA[\mathrm{m}]$, $\LmatPhi[\mathrm{m}]$, $\matr{Z}$ and $\matr{Y}_{\mathrm{eq}}$. A computationally efficient method for solving~\eqref{eq:AEFIEoutJ} is to use an iterative solver with the constraint preconditioner in~\cite{Qian2009}. This method only requires the ability to multiply a vector by the system matrix and does not require assembling matrices explicitly.
\section{Proposed Parallelization}\label{sec:prop-para}%
In this section, we propose an efficient parallelization strategy for solving the system of equations in~\eqref{eq:AEFIEoutJ} on a distributed-memory computer. To maximize parallel scalability, we treat all matrices in~\eqref{eq:AEFIEoutJ} as parallel distributed matrices, including all $\matr{Z}_i$ and $\matr{Y}_{i,\mathrm{eq}}$.
 
We choose two cooperative methods of workload decomposition, where one is targeted at the external workload and the other is optimized for the workloads of the internal problem. The two methods for workload decomposition result in compatible distributions of $\Hmat$ and can be linked together through a data redistribution step with minimal communication complexity.

\subsection{Analysis of the Computational Workload}\label{sec:para-analysis}%
The computational complexity of constructing the system matrix or multiplying a vector by the system matrix is dominated by three computations. The first pertains to the external problem, involving $\LmatA[\mathrm{m}]$ and $\LmatPhi[\mathrm{m}]$. The second involves the internal problem in the equivalent configuration and requires the inversion of $\matr{L}_{l}$ as in~\eqref{eq:EFIEeqYeq}. The third involves the internal problem, requiring the inversion of $\Lmat$ as in~\eqref{eq:MFIEinZin}. Due to their differences, these computations use the following three different approaches:
\begin{enumerate}
	\item The AIM can be employed to rapidly compute matrix-vector products involving $\LmatA[\mathrm{m}]$ and $\LmatPhi[\mathrm{m}]$~\cite{Bleszynski1996}. In the AIM, matrices $\LmatA[\mathrm{m}]$ and $\LmatPhi[\mathrm{m}]$ are decomposed as
	\begin{align}
	\matr{L}_{\mathrm{NR}} + \matr{W}\matr{H}\matr{P},\label{eq:AIM_L}
	\end{align}
	where $\matr{L}_{\mathrm{NR}}$ is a sparse matrix that models near-region interactions with entries computed by numerical integrals and the second term is a set of operations that quickly computes far-region interactions~\cite{Bleszynski1996}. This quick computation is achieved as follows: matrix $\matr{P}$ projects mesh-based sources onto an auxiliary uniform 3-D grid; grid potentials due to grid sources are computed through fast Fourier transforms (FFT)s, which are responsible for the acceleration of the method, and are represented by the matrix $\matr{H}$; and finally, the interpolation matrix $\matr{W}$ maps grid potentials back onto mesh-based testing functions. Since FFTs are a global operation, $\matr{L}_{\mathrm{NR}}$ is precorrected to prevent double-counting the near-region terms~\cite{Phillips1997}. Thus, the computational workload associated with the external problem, i.e., constructing and computing matrix-vector products with the matrices $\LmatA[\mathrm{m}]$ and $\LmatPhi[\mathrm{m}]$ is a large single parallel task.
	
	\item In many applications, such as metamaterials and electrical interconnects, each object has a small number of unknowns when compared to the total number of unknowns in the system of equations. Thus, it is reasonable to implement $\matr{L}_{l}$ and $\matr{K}_{l}$ for each object as dense matrices. Then, $\left(\matr{L}_{l}\right)^{-1}$ is implemented as a dense LU factorization followed by forward and backward substitutions. The LU factorization is completed in a preprocessing step before the iterative solver begins. In the case of large individual objects, the AIM can also be implemented in the internal problem~\cite{Sharma2020_aim}. Unlike the external problem, the internal problem is made up of numerous independent tasks of varying sizes that involve constructing and performing operations with smaller distributed matrices.
	
	\item In general, the objects are lossy conductors which at high frequency will have small skin depths. Thus, an effective optimization is to treat $\Lmat$ and $\Kmat$ as sparse matrices by ignoring physically distant interactions~\cite{Qian2007}. As a result, multiplying a vector by $\left(\Lmat\right)^{-1}$ can be accomplished by a sparse direct solver. At low frequency, this condition is violated because the skin depth may be larger than the dimensions of the objects, so $\Lmat$ and $\Kmat$ must be implemented as dense matrices or accelerated with the AIM~\cite{Sharma2020_aim}. If the high frequency optimization is used, then this workload will be considerably smaller than the workload described in point 2) and can be ignored for the purpose of workload balancing. If the optimization cannot be used, the workload of this step is identical to the workload described in point 2), and a single balancing strategy can be used for both workloads.
\end{enumerate}

\subsection{Parallelization of the External Problem}\label{sec:para-ext}%
If we do not consider the coupling to the internal problem, the workload of the external problem is analogous to the workload that arises from formulations that only consider PEC objects or make use of the SIBC~\cite{Yuferev2009}. Hence, we employ the highly-efficient parallelization strategy proposed in~\cite{Marek2020b}. In~\cite{Marek2020b}, graph partitioning algorithms are used to distribute near-region computations equally over the processes, while minimizing the amount of communication required in matrix-vector products involving $\LmatA[\mathrm{m}]$ and $\LmatPhi[\mathrm{m}]$. Furthermore, FFTs are needed in the AIM for computing far-region interactions. These operations are most efficiently parallelized by the \mbox{2-D} pencil decomposition proposed in~\cite{Wei2014} for homogeneous media and in~\cite{Liu2019} for multilayered media. These \mbox{2-D} pencil decompositions are adopted in the proposed method.

The graph partitioning algorithm favors solutions that minimize the number of partitions per object, and it ensures that the mesh of a sufficiently small object is not unnecessarily partitioned. Using graph partitioning in the external problem has the added benefit of producing mesh partitions that are compatible with the workload balancing strategy used in the internal problem, which will be introduced in the next section.

\subsection{Parallelization of the Internal Problem}\label{sec:para-int}%
The operations of the internal problem are generating $\matr{Y}_{\mathrm{eq}}$ and $\matr{Z}$ and multiplying $\Hmat$ by them, as in~\eqref{eq:AEFIEoutJ} and~\eqref{eq:Cdef}. Naturally, computing and factorizing all $\matr{Y}_{i,\mathrm{eq}}$ and $\matr{Z}_i$ is a set of independent parallel tasks. Provided that the unknowns of $\Hmat$ can be redistributed efficiently to processors, which will be discussed in~\secref{sec:para-redist}, then multiplying $\Hmat$ by $\matr{Y}_{\mathrm{eq}}$ or $\matr{Z}$ is also a set of $N_\mathrm{obj}$ independent parallel tasks.

One parallelization approach is to perform the operations of the internal problem with processes that are assigned to the corresponding mesh partitions in the external problem. This approach is called the ``No Redistribution'' strategy, since it results in very minor communication during the data redistribution stage. However, as numerical results will show, the ``No Redistribution'' strategy performs poorly, since it ignores the differences between the internal and external workloads, so it fails to balance the internal problem optimally. Furthermore, the critical issue with the ``No Redistribution'' strategy is that it leads to situations where a task cannot be started because some of the processors in the task's group are still completing an earlier task. 

In the proposed method, we recognize that balancing the internal workloads is a task scheduling problem~\cite{Drozdowski2009}, where each task consists of all the internal problem computations associated with a single conductive object. By carefully assigning tasks to processors using algorithms from scheduling theory, the proposed method entirely avoids the critical issue of the ``No Redistribution'' strategy. 

\subsubsection{The Longest Processing Time Algorithm}\label{sec:para-int-sched}%
Scheduling theory is a vast research topic that covers a range of differing problem types. For clarity, we first introduce notation and concepts for the simpler problem of distributing a set of tasks $T = \left\{T_1, \dots, T_{N_{T}} \right\}$ over $P$ identical processing cores, with the restriction that each task will be executed by only one processor core. Later, we discuss the problem of distributing a set of parallel tasks, where multiple processor cores can perform a single task. 

Specifically, we seek a schedule $C$ which assigns tasks in $T$ to individual processors with the criterion of minimizing the duration of $C$. The schedule $C$ can be represented by a 2-D array with entries $c_{p,t}$, where $c_{p,t}$ is the $t$-th task assigned to processor $p$ and each processor is assigned $N_{\mathrm{T},p}$ total tasks.

The length of $C$ is defined as the maximum time a processor takes to finish its list of tasks, $C_{\mathrm{max}} = \max_p{F_{p}}$, where the finish time $F_{p}$ for processor $p$ is given by
\begin{align}
F_{p} = \sum^{N_{\mathrm{T},p}}_{t=1}{\tau\left(T_{c_{p,t}}\right)},\label{eq:F_comptime}
\end{align}
with $\tau$ being a function that returns the execution time of a given task~\cite{Belkhale1990}.

The problem of finding a schedule $C$ with minimal length can be solved approximately by the longest processing time (LPT) algorithm~\cite{Graham1969}. The LPT algorithm first sorts the tasks in $T$ in order of descending execution time. Then, the algorithm iterates through $T$, and assigns $T_i$ to the processor with the earliest finish time. The LPT algorithm is presented in detail as Algorithm~\ref{fig:alg-lpt}.

The LPT algorithm is a greedy method that is guaranteed to produce a schedule no longer than $R \cdot C^{\mathrm{opt}}_{\mathrm{max}}$, where $C^{\mathrm{opt}}_{\mathrm{max}}$ is the length of the optimal schedule and $R$ is the worst-case ratio~\cite{Graham1969}
\begin{align}
R = \frac{4}{3} - \frac{1}{3P}.\label{eq:R_LPT}
\end{align}
\begin{figure}[!t]
	\setlength{\intextsep}{0pt} 
	\removelatexerror
	\begin{algorithm}[H]
		\caption{Longest Processing Time (LPT)}
		\label{fig:alg-lpt}
		\begin{algorithmic}
			\Require List $T$ sorted in descending order with respect to $\tau$
			\For {$i = 1$ to $N_\mathrm{T}$}
			\State Find processor $p$ with earliest $F_p$
			\State Append $T_i$ to the $p$-th row of $C$ 
			\State $F_p = F_p + \tau\left(T_i\right)$
			\EndFor 
			\State \textbf{return} $C$ 
		\end{algorithmic}
	\end{algorithm}
\end{figure}

Since several electromagnetic problems involve large objects, it is desirable to be able to assign multiple processing cores to some tasks. In this case, we require an algorithm that solves the more difficult problem of parallel task scheduling (PTS). In a PTS problem, each $T_i$ is called a moldable task that may be executed with $P_i>1$ processing cores~\cite{Drozdowski2009}. As a result, solving a PTS problem  requires an algorithm to choose the number of processors $P_i$ in addition to building a schedule $C$. This type of problem can be solved approximately by the Part\_Schedule algorithm~\cite{Belkhale1990}.

The Part\_Schedule algorithm builds a schedule through iterations. To begin, all $P_i$ are set to 1, and a schedule is constructed using the LPT algorithm. In each subsequent iteration, the task $T_i$ with longest duration $\tau\left(T_i\right)$ is selected and its corresponding $P_i$ is incremented. Then, a new schedule is built using the LPT algorithm. However, the LPT algorithm must be modified because now there is a mixture of sequential and parallel tasks. In addition, to eliminate the possibility of idle processors in between tasks, all processors assigned to a parallel task must begin execution of the task at the same time. These two issues are remedied by assigning all parallel tasks first and requiring that all parallel tasks can be executed simultaneously. In other words, the condition $\sum_{P_i>1} P_i \leq  P$ must be satisfied~\cite{Belkhale1990}. After parallel tasks have been added to $C$, the remaining sequential tasks are assigned according to the LPT algorithm. The algorithm terminates whenever a new schedule is built that is longer than the current schedule, or if $P_i$ cannot be incremented further because the condition $\sum_{P_i>1} P_i \leq  P$ would be violated. The Part\_Schedule algorithm can be proved to find a schedule with worst-case ratio~\cite{Belkhale1990}
\begin{align}
R = \frac{2}{1-\frac{1}{P}}.\label{eq:R_part}
\end{align}

\subsubsection{Proposed Workload Balancing Algorithm}\label{sec:para-int-algo}%
We propose to use the Part\_Schedule algorithm to balance the internal workloads. Specifically, we seek a schedule of minimal length where the workload of the $i$-th object is represented in the schedule by the task $T_i$. Since the duration of a task is not known a priori, we assume that the duration $\tau\left(T_i\right)$ is proportional to the number of operations required by a workload. Thus, we associate $T_i$ with an estimated workload $W_i$, which is dominated by dense matrix operations involving $\matr{Y}_{i,\mathrm{eq}}$.

The workload associated with generating $\matr{Y}_{i,\mathrm{eq}}$ requires $\mathcal{O}(n^2)$ operations to compute matrix entries and $\mathcal{O}(n^3)$ operations to compute the LU factorization. In addition, multiplying a vector by $\matr{Y}_{i,\mathrm{eq}}$ requires $\mathcal{O}(n^2)$ operations to compute the matrix-vector product $\left(\matr{K}_{l} + \frac{1}{2}\matr{I}_{\mathrm{x}} \right) \Hmat$ and perform the forward and backward substitutions associated with $\left(\matr{L}_{l}\right)^{-1}$. Although LU decomposition has, in theory, a greater computational complexity, we use $\mathcal{O}(n^2)$ as a workload estimate for two reasons. First, we assume that the objects are not extremely large, so other matrix operations have comparable complexity. Second, on present-day computer hardware dense matrix factorization is usually constrained by memory, which also scales by $\mathcal{O}(n^2)$~\cite{Dongarra1994}. For these reasons, we estimate the workload as $W_i = N_{\mathrm{E},i}^2$, where $N_{\mathrm{E},i}$ is the number of mesh edges belonging to the $i$th object. The proposed workload estimate comes with the added benefit that memory usage will also be balanced among processors.

We allow each $T_i$ to be executed with $P_i$ processes, which results in an estimated task duration time of $\tau\left(T_i\right) = W_i/ P_i$. For simplicity, we assume that $P_i$ is small enough that the parallel efficiency of $T_i$ remains high. However, this estimate could be replaced with a more accurate relationship that accounts for decreasing parallel efficiency of the task for large $P_i$. The proposed method would otherwise remain unchanged.

\begin{figure}[!t]
	\centering
	\subfloat{\includegraphics[width=3.45in]{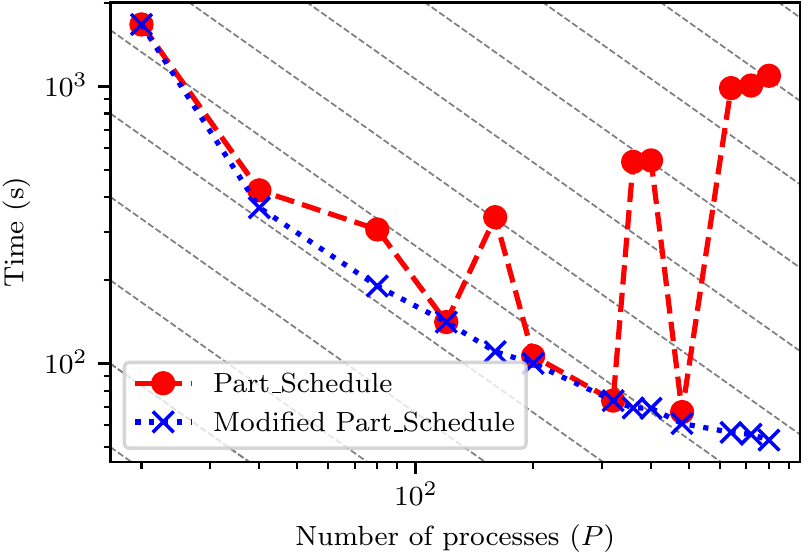}}%
	\caption{Time spent computing the complete system matrix-vector product for the fan-out interposer in~\secref{sec:fan-out}. Times are reported for a solver using the Part\_Schedule and modified Part\_Schedule algorithms. Gray dashed lines are ideal scalability slopes.}
	\label{fig:times-approx-square}
\end{figure}

The proposed method leads to a PTS problem that can be solved by the Part\_Schedule algorithm~\cite{Belkhale1990}. However, the original Part\_Schedule algorithm proposed in~\cite{Belkhale1990} leads to fluctuations in parallel efficiency as $P$ increases. Figure~\ref{fig:times-approx-square} plots the overall time spent computing matrix-vector products for the fan-out interposer in~\secref{sec:fan-out} when using the original Part\_Schedule algorithm. For some values of $P$, the parallel efficiency is high, but for others it is quite low. The low efficiencies are due to the Part\_Schedule algorithm assigning arbitrary values to $P_i$ that do not allow for efficient distributed dense matrices. For example, the \mbox{2-D} block cyclic decomposition used in the distributed-memory linear algebra library ScaLAPACK~\cite{slug} is most efficient when the number of processors can be factored into roughly equal numbers, since the communication complexity is proportional to the sum of the factors~\cite{Dongarra1994}. Therefore, we propose to use a modified Part\_Schedule algorithm that restricts values of $P_i$ to be approximate squares, i.e., $P_i = Q^2$ or $P_i = Q(Q + 1)$ where $Q \in \mathbb{N}$. If $P_i$ is sufficiently small, the communication costs will also be small, so this restriction is unnecessary. Thus, when $P_i \leq P_{\mathrm{cutoff}}$, this restriction is removed, which allows for the workload to be more evenly balanced. Figure~\ref{fig:times-approx-square} shows that the modified Part\_Schedule algorithm results in a consistent and monotonic decrease in CPU times as the number of processes is increased. For all tests that use the proposed method, we set ${P_{\mathrm{cutoff}}=20}$. This value for $P_{\mathrm{cutoff}}$ is chosen because for any ${P_i>20}$, the closest approximate square is within a $20\%$ variation from  $P_i$. This range is adequate for the purpose of workload balancing. A detailed description of the modified Part\_Schedule algorithm is presented in Algorithm~\ref{fig:alg-part}.

\begin{figure}[!t]
	\setlength{\intextsep}{0pt} 
	\removelatexerror
	\begin{algorithm}[H]
		\caption{Modified Part\_Schedule}
		\label{fig:alg-part}
		\begin{algorithmic}
			\Require List $T$ sorted in descending order with respect to $\tau$
			\For {$i = 1$ to $N_\mathrm{T}$}
			\State $P_i = 1$
			\EndFor
			\State Let $a = P$
			\State $C \gets \mathrm{LPT}(T)$ \Comment{Initialize $C$ using the LPT algorithm.}
			\While {$a > 0$}
			\State Find $T_i$ with max $\tau\left(T_i\right)$
			\State Let $h =\tau\left(T_i\right)$
			\If {$P_i < P_{\mathrm{cutoff}}$}
			\State Let $d = 1$
			\Else
			\State Let $d = \mathrm{NextApproximateSquare}(P_i)$
			\EndIf
			\If {$P_i == 1$}
			\State $a = a - d - 1$
			\Else
			\State $a = a - d$
			\EndIf
			\If {$C_{\mathrm{max}} \neq h$ or $a < 0$}
			\State \textbf{break}
			\EndIf
			\State $P_i = P_i + d$
			\State $C^\prime \gets \mathrm{LPT}(T)$
			\If {$C^\prime_{\mathrm{max}} \ge C_{\mathrm{max}}$}
			\State $P_i = P_i - d$  \Comment{Undo increment.}
			\State \textbf{break}
			\Else
			\State $C \gets C^\prime$
			\EndIf
			\EndWhile
			\State \textbf{return} $C$ 
		\end{algorithmic}
	\end{algorithm}
\end{figure}

\begin{figure}[!t]
	\centering
	\subfloat{\includegraphics[width=3.49in]{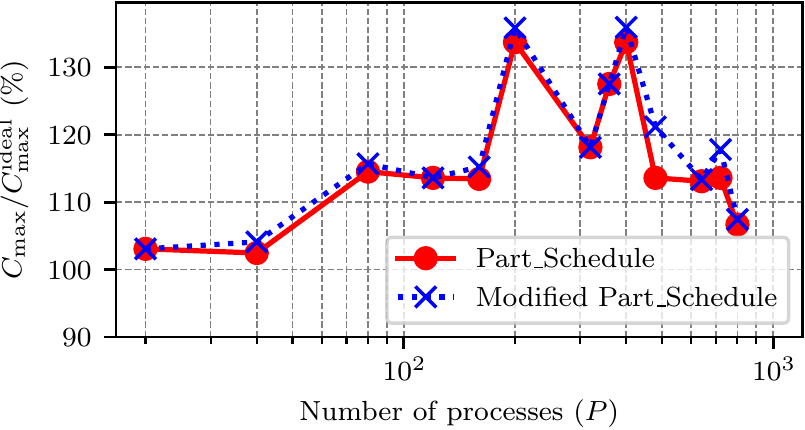}}%
	\\
	\centering
	\subfloat{\includegraphics[width=3.49in]{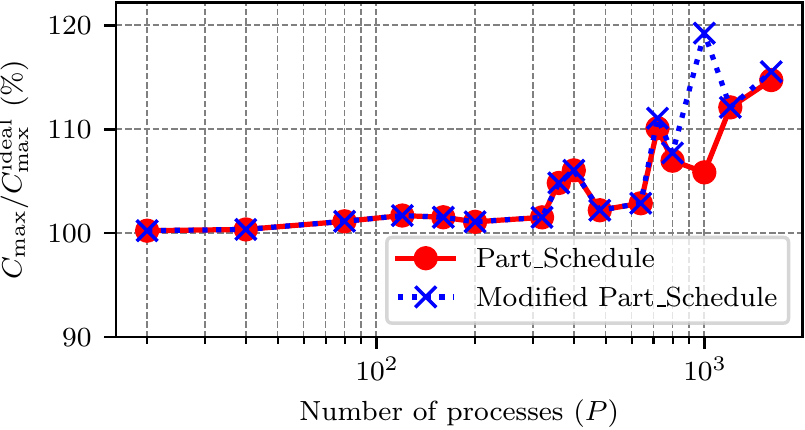}}%
	\caption{Schedule lengths resulting from the Part\_Schedule and the modified Part\_Schedule algorithms for structures in~\secref{sec:val-scala}. Top panel: Schedules built for the interposer structure. Bottom panel: Schedules built for the split-ring resonator array.}
	\label{fig:sched-lengths}
\end{figure}

The modified Part\_Schedule algorithm seems to impose harsh restrictions that could impact the quality of the constructed schedules. However, we demonstrate in~\figref{fig:sched-lengths} that schedule lengths resulting from the modified Part\_Schedule algorithm are comparable to those generated by the original algorithm. Notice that the schedule lengths reported in~\figref{fig:sched-lengths} are based on workload estimates and not based on execution times. Since the optimal schedule cannot be easily found, we compare the schedule lengths produced by the two algorithms by dividing them by the length of an ideal schedule $C^{\mathrm{ideal}}_{\mathrm{max}}$. The ideal schedule length is calculated by assuming that the total workload of the internal problem can be balanced perfectly. In fact, this ideal schedule is likely not realizable and $ C^{\mathrm{opt}}_{\mathrm{max}} \geq C^{\mathrm{ideal}}_{\mathrm{max}}$. For most values of $P$, the schedules generated by the two algorithms are identical in length, and the largest difference between two lengths is less than 15\%.

\subsection{Data Redistribution Stage}\label{sec:para-redist}%
The entries of $\Hmat$ are needed by potentially different processes on different compute nodes in the external and in the internal workloads. As result, a communication stage is needed at each solver iteration to redistribute some of the entries of $\Hmat$ in the external problem to a form suitable for the internal problem. In the proposed method, the data is redistributed through the VecScatter functionality provided by PETSc~\cite{petsc-efficient}. This leads to every process potentially communicating with every other process, which at worst would require $\mathcal{O}\left(P^2\right)$ messages.

We reduce the number of messages in this communication stage by allowing each process to choose a list of tasks in $C$, i.e., a row of $C$, that contains the objects with the greatest number of mesh edges in common with the process' mesh partition of the external problem. Specifically, the following steps are used:

\begin{enumerate}
\item Each process counts $N_{\mathrm{E}, p}$, the number of mesh edges belonging to objects in the $p$-th list of tasks that are also present in the processes' mesh partition of the external problem;
\item Each process ranks all task lists in decreasing order of $N_{\mathrm{E},p}$;
\item Finally, starting from the lowest rank, each process is assigned a task list, from those that remain, with the highest $N_{\mathrm{E},p}$.
\end{enumerate}

The mesh partitions created for the external problem usually only include a single object, this is especially true when $P$ is larger than the number of objects. Consequently, the data redistribution method works extremely well in practice. This stage takes a negligible amount of time in the scalability experiments run in~\secref{sec:val-scala}.

As a result, both the external and internal problems are parallelized with high efficiency, while communication costs are kept small. In~\secref{sec:val-scala}, we demonstrate an efficient electromagnetic solver based on the proposed method that rigorously models the physics both inside and outside each conductive object.

\subsection{Implementation Details}\label{sec:para-impl}%
The proposed implementation uses the GMRES algorithm~\cite{Saad1986} to solve the system of equations in~\eqref{eq:AEFIEoutJ}. The PETSc library~\cite{petsc-efficient} is used as an overall framework for distributed sparse matrix and vector operations, and PETSc partitions matrices by distributing contiguous blocks of rows to processes. Second, all operations involving distributed dense matrices, as in point 2) of~\secref{sec:para-analysis}, are performed by ScaLAPACK~\cite{slug}, which uses a 2-D block-cyclic distribution. Finally, the direct solution of large sparse linear systems, as in point 3) of~\secref{sec:para-analysis}, is provided by the MUMPS library~\cite{MUMPS:1}.

In the proposed method, small independent groups of processors work on sparse and dense matrix operations needed for the internal problem. These groups of processors are implemented using sub-communicators provided by the message passing interface (MPI). There is an MPI communicator associated with each object in the mesh, and thus to each pair of matrices $\matr{Z}_i$ and $\matr{Y}_{i,\mathrm{eq}}$. In addition, there is an MPI world communicator that is associated with all external-problem matrices. These MPI communicators allow for communication and synchronization among only the MPI processes that are required for a specific operation.%
\section{Numerical Results}\label{sec:val-scala}
\begin{figure}[!t]
	\centering
	\includegraphics[width=3.49in]{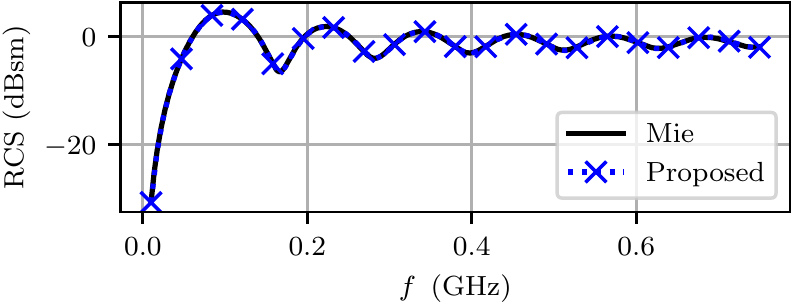}%
	\caption{Monostatic RCS of a 1-m diameter copper sphere computed with the Mie series and the proposed solver.}
	\label{fig:sphereRCS}
\end{figure}

In this section, we validate the accuracy of a parallel solver based on the proposed parallelization strategy and report its weak and strong scalability performance. All simulations were executed on the SciNet Niagara supercomputer~\cite{Ponce2019}, which consists of compute nodes with 40 Intel Skylake cores operating at 2.4 GHz. In addition, each node has 202 GB of available memory, and internodal communication is supported by an Infiniband network with a Dragonfly topology.

The Intel Skylake cores in Niagara have the Intel Turbo Boost feature enabled that dynamically changes the clock speed based on the number of cores currently active on the CPU. This feature confounds the evaluation of parallel scalability. Thus, we chose to start parallel scalability experiments at $P=20$ and fully saturate each CPU socket with processes to avoid this issue.

\begin{figure}[!t]
 \centering
 \includegraphics[width=3.49in]{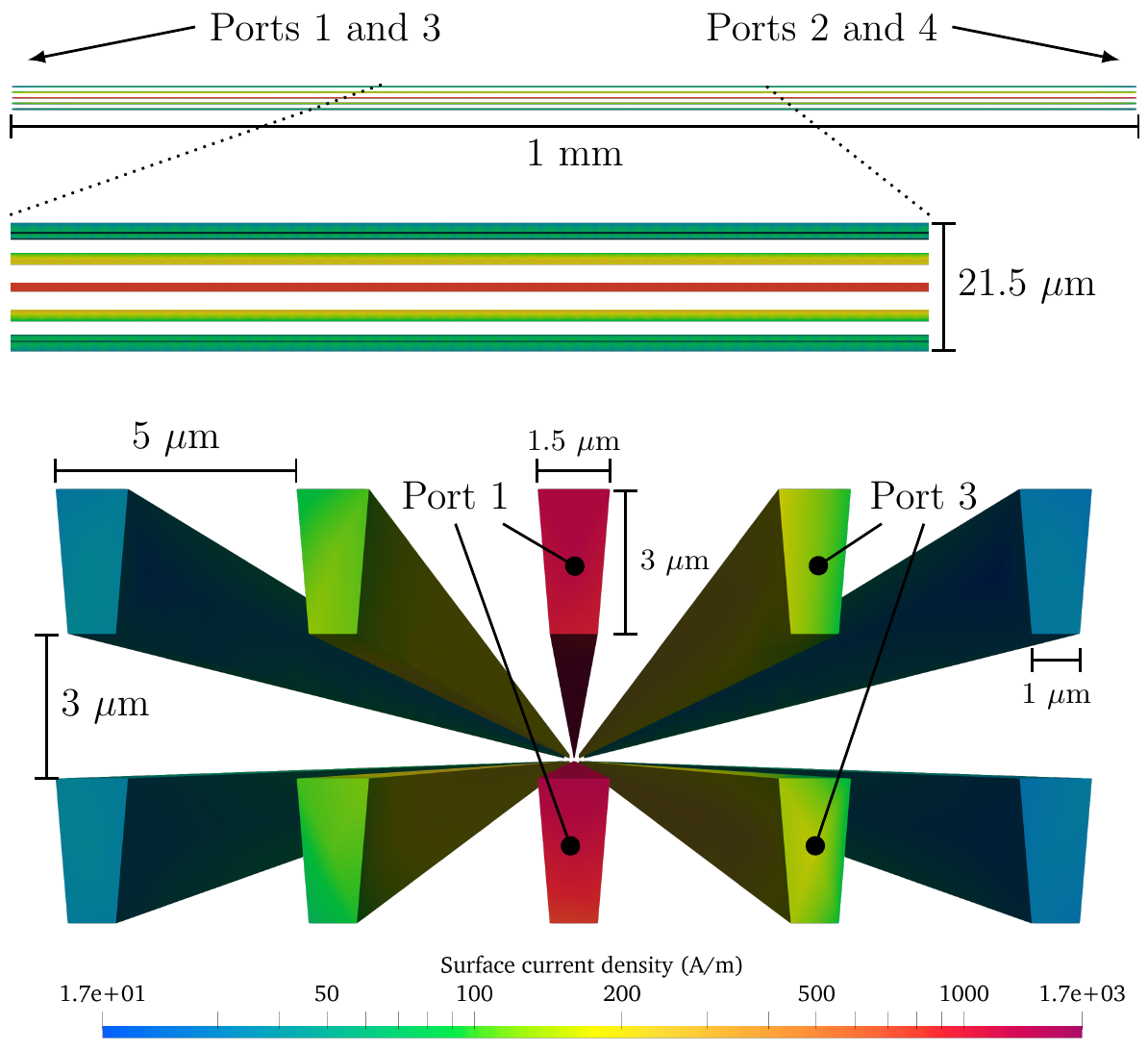}%
 \caption{Geometry of the high-speed bus with 5 pairs of trapezoidal conductors. The induced electric current density due to a 10 GHz excitation at port 1 is also displayed.}
 \label{fig:traps-geom}
 \end{figure}
\begin{table}[!t]
	\renewcommand{\arraystretch}{1.3}
	\caption{Dimensions and material properties of the background layered media used for electrical interconnect simulations.}
	\label{tbl:laymed}
	\centering
	\begin{tabular}{|c|c|c||c|c|c|}
		\multicolumn{3}{c}{High-Speed Bus}&\multicolumn{3}{c}{Fan-Out Interposer}\\
		\multicolumn{3}{c}{(\secref{sec:trap-prism})}&\multicolumn{3}{c}{(\secref{sec:fan-out})}\\
		\hline
		$h$ ($\mu$m) & $\epsilon_r$ & $\sigma$ (S/m) & $h$ ($\mu$m) & $\epsilon_r$ & $\sigma$ (S/m)\\
		\hhline{|=|==#=|==|}
		$\infty$ & \multicolumn{2}{c||}{Air} &$\infty$ & \multicolumn{2}{c|}{Air}\\
		\hline
		$30$ & $4.0$ & $0$ &$27.5$ & $4.0$ & $0$ \\
		\hline
		$495$ & $11.9$ & $10$ &$497.5$ & $11.9$ & $10$ \\
		\hline
		$\infty$ & \multicolumn{2}{c||}{PEC} & $\infty$ & \multicolumn{2}{c|}{PEC}\\
		\hline
	\end{tabular}
\end{table}

\subsection{Copper Sphere}\label{sec:sphere}
First, the proposed solver's accuracy is validated for electromagnetic scattering problems in homogeneous media with a field excitation. The monostatic radar cross section (RCS) of a 1-m diameter copper sphere is computed using a mesh composed of 2,934 edges and 1,956 triangles. The dimensions of the AIM grid are $20 \times 20 \times 20$. Results from the proposed solver are compared to data obtained from a Mie series solution in~\figref{fig:sphereRCS}, and both results are in excellent agreement.

\subsection{High-Speed Bus with Trapezoidal Conductors}\label{sec:trap-prism}
Next, the proposed solver is validated for conductors embedded in a multilayered medium with port excitation through a trapezoidal conductor structure. The trapezoidal cross-section represents the result of imperfect fabrication. 

The structure of the high-speed bus is displayed in~\figref{fig:traps-geom} and is made up of identical $1$-mm long copper trapezoidal prisms with $\sigma=5.8\times 10^7$ S/m. The conductors are placed every $5~\mu$m horizontally and spaced $3~\mu$m apart in the vertical direction. The total width of the high-speed bus is $21.5~\mu$m. The entire structure is embedded in a multilayered medium described in the first part of~\tblref{tbl:laymed}. A total of four lumped ports are used to excite the interconnect and are placed at both ends.

\begin{figure}[!t]
	\centering
	\subfloat{\includegraphics[width=3.49in]{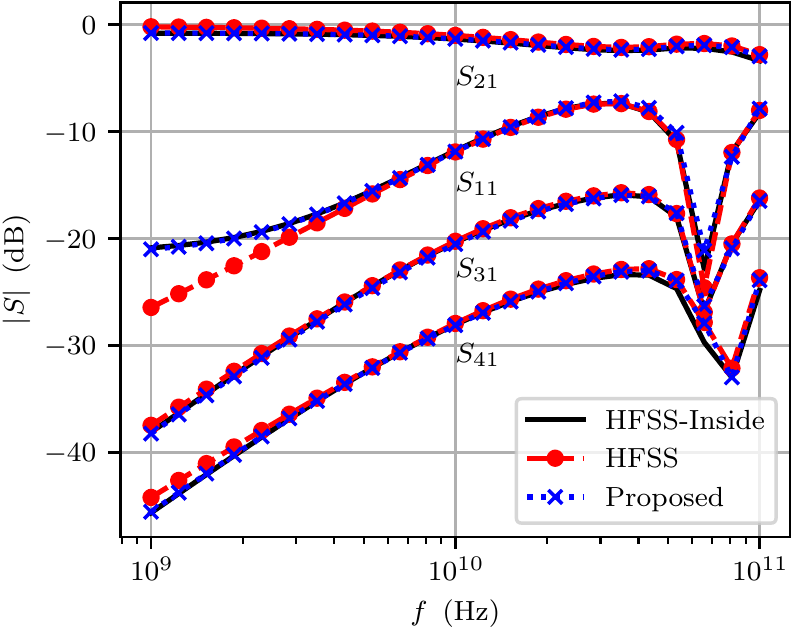}}%
	\hfill
	\subfloat{\includegraphics[width=3.49in]{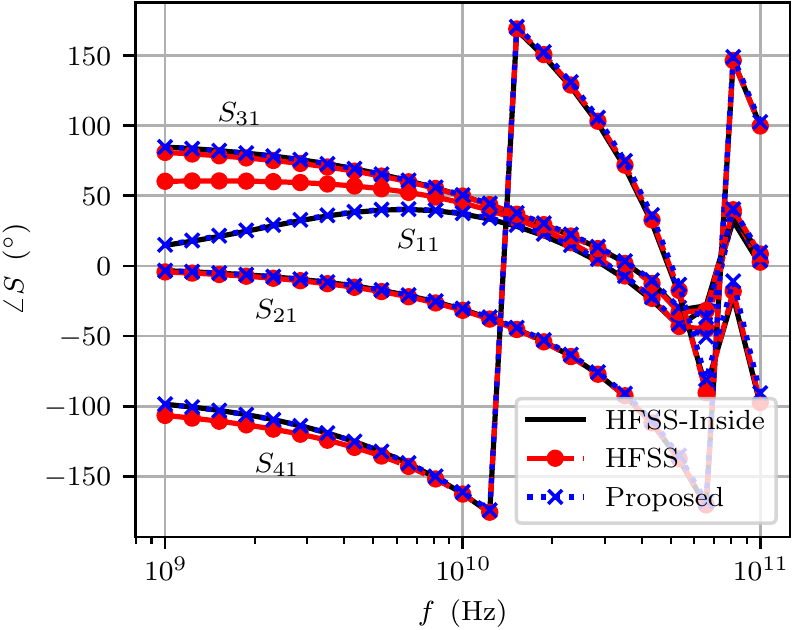}}%
    \caption{Computed scattering parameters for the high-speed bus with 5 pairs of conductors using HFSS and the proposed solver. Top panel: Magnitude of the scattering parameters. Bottom panel: Phase of scattering parameters.}
	\label{fig:s-trapezoids}
\end{figure}

The accuracy of the proposed solver is confirmed over frequencies ranging from 1 GHz to 100 GHz by comparing computed scattering parameters to those calculated by a commercial tool (Ansys HFSS 2020 R2). Two sets of results were obtained from HFSS, the first named ``HFSS-Inside'' was calculated by enabling the ``Solve Inside'' option in the HFSS user interface. This forces HFSS to mesh the interior of conductive objects and leads to a more accurate solution at low frequencies. The other set of results denoted as ``HFSS'' were obtained by turning off this feature, in which case HFSS will use an approximate model for skin effect. These results are reported in~\figref{fig:s-trapezoids} and excellent agreement is observed between the proposed method and the ``HFSS-Inside'' curves. At high frequency, all three methods are in excellent agreement. However, at low frequencies the HFSS solution based on an approximate model is inaccurate, underlining the importance of using an accurate model for skin effect which resolves fields inside conductive objects.

\begin{table}[!t]
	\renewcommand{\arraystretch}{1.3}
	\caption{Parameters for high-speed bus structures.}
	\label{tbl:trap-structures}
	\centering
	\begin{tabular}{|c||c|c|c|c|}
		\hline
		Structure & $P$ & $N_x \times N_y \times N_z$&$N_{\mathrm{E}}$&$N_{\mathrm{iter}}$\\
		\hhline{|=#=|=|=|=|}
		5 Pairs & $20$ & $500 \times 20 \times 8$ & $70,260$ & $66$\\
		\hline
		10 Pairs & $40$ & $500 \times 40 \times 8$ & $140,520$& $74$\\
		\hline
		20 Pairs & $80$ & $500 \times 80 \times 8$ & $281,040$& $85$\\
		\hline
		40 Pairs & $160$ & $500 \times 160 \times 8$ & $562,080$& $96$\\
		\hline
		80 Pairs & $320$ & $500 \times 320 \times 8$ & $1,124,160$& $109$\\
		\hline
		160 Pairs & $640$ & $500 \times 640 \times 8$ & $2,248,320$& $124$\\
		\hline
		250 Pairs & $1000$ & $500 \times 1000 \times 8$ & $3,513,000$& $135$\\
		\hline
	\end{tabular}
\end{table}
 
Next, the high-speed bus is used to test the weak scalability of the proposed solver. In a weak scalability experiment, program execution times are recorded as the number of processes is increased, while keeping the computational workload per process approximately constant. To this end, the number of trapezoidal conductors is increased proportionally to the number of processes used from 5 pairs up to 250 pairs. The extra trapezoidal conductors are added on either side of the prisms shown in~\figref{fig:traps-geom}, so that the total length of the structure remains unchanged. The number of port excitations is also kept constant. Table~\ref{tbl:trap-structures} lists parameters for the additional structures that were tested, where $N_{\mathrm{iter}}$ is the number of iterations required in the GMRES algorithm~\cite{Saad1986} to reach a residual norm with relative tolerance of $10^{-6}$. In addition,~\tblref{tbl:trap-structures} shows the number of MPI processes $P$ used. In the proposed method, these choices of $P$ always lead to the assignment of two processes per conductor. 
 \begin{figure}[!t]
 	\centering
 	\subfloat{\includegraphics[width=3.49in]{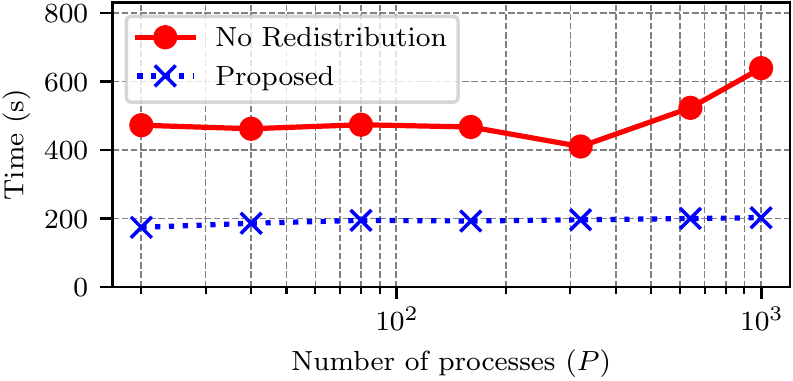}}%
 	\\
 	\centering
 	\subfloat{\includegraphics[width=3.49in]{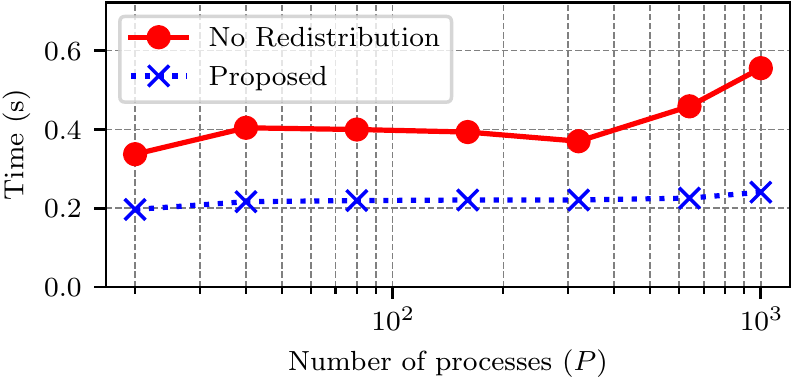}}%
 	\\
 	\centering
 	\subfloat{\includegraphics[width=3.49in]{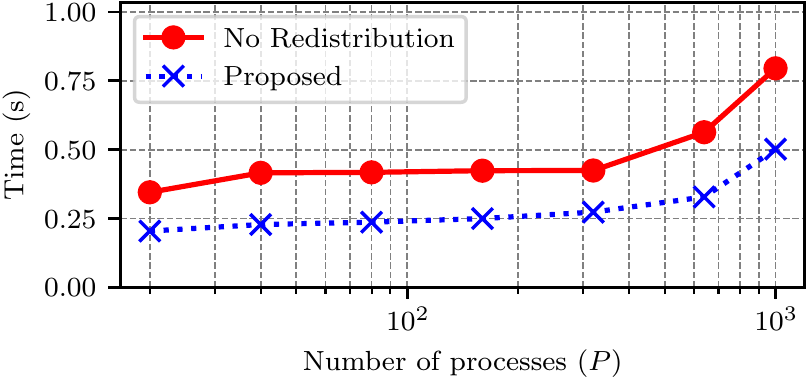}}%
 	\caption{Execution times for the high-speed bus in~\secref{sec:trap-prism} using the proposed solver and the ``No Redistribution'' method. Top panel: matrix generation. Center panel: average time to compute a matrix-vector product. Bottom panel: average time to perform one iteration of the GMRES algorithm~\cite{Saad1986}.}
 	\label{fig:scala-trapezoids}
 \end{figure}

The results of the weak scalability experiment are presented in~\figref{fig:scala-trapezoids}, where the proposed method and ``No Redistribution'' methods for parallelization are compared. The plots in~\figref{fig:scala-trapezoids} clearly show the superiority of the proposed method for all major stages of the solver, i.e., the matrix generation and system solution stages. The center panel of~\figref{fig:scala-trapezoids} reports the average time to compute the system matrix-vector product and excludes the preconditioning stage.

The performance of the ``No Redistribution'' method is inferior, because it leads to the situation where many processors remain idle until previous work is finished. For example, suppose processor two is assigned to trapezoidal prism A and B, processor one is assigned to prism A, and processor three is assigned to prism B. While processor one and two perform work on prism A, processor three must remain idle and cannot start its computations on prism B until processor two is ready. Consequently, many processors in the ``No Redistribution'' method are blocked increasing execution times.

The weak scalability plots also vindicate the choice of workload scaling as the proposed method's execution times remain constant until large $P$. However, the bottom panel of~\figref{fig:scala-trapezoids} indicates that the time to solve the system of equations begins to grow for $P$ greater than 400. This increased time is due to decreasing parallel efficiency in MUMPS~\cite{MUMPS:1} during the preconditioning stage and is not related to workload balancing issues in the proposed method.
\begin{figure}[!t]
	\centering
	\includegraphics[width=3.49in]{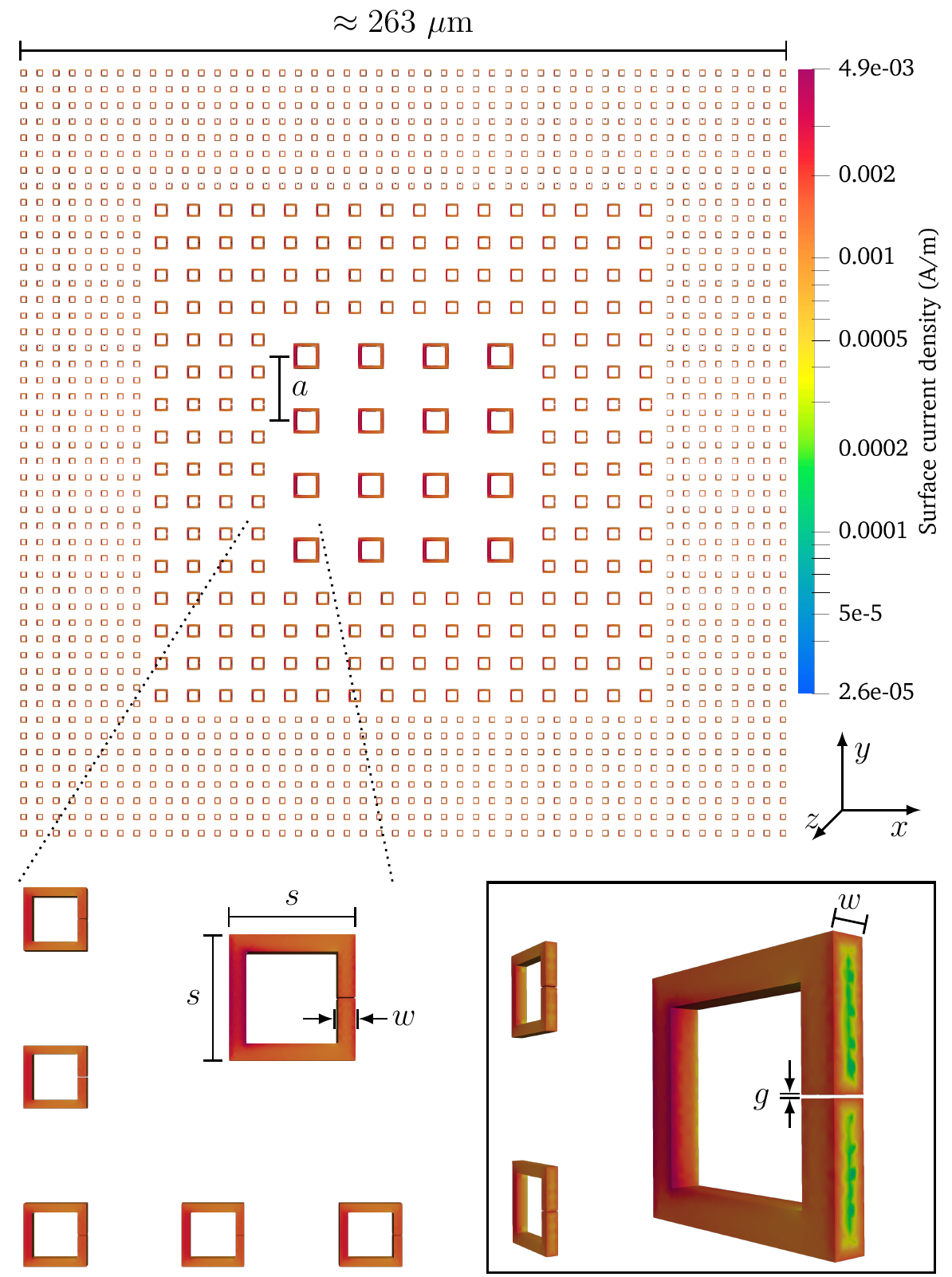}%
	\caption{Geometry and electric current density of the SRR array in~\secref{sec:srr}.}
	\label{fig:srr-geom}
\end{figure}
\subsection{Split Ring Resonator Array}\label{sec:srr}
The strong scalability of the proposed method is tested on an example related to metamaterials~\cite{Guney2009}. The metamaterial structure under study, illustrated  in~\figref{fig:srr-geom}, is an array of split-ring resonators in free space with elements of varying sizes.

There are three versions of the SRR element in the array that are each characterized by their size $s$, width $w$, gap $g$, and unit-cell size $a$. These SRR elements are similar to those studied in~\cite{Guney2009}, except these are made of copper with $\sigma=5.8\times 10^7$ S/m. The largest SRR element in the array has the following dimensions: $s=8.82~\mu$m, $w=1.2~\mu$m, $g=0.1~\mu$m, and $a=22.2~\mu$m.  In addition, the dimensions of smaller elements in the center and outer ring of~\figref{fig:srr-geom} are the same as the largest SRR element but scaled down by a factor of 2 and 4, respectively. In total, there are 1,488 objects in the array, 1,281,975 mesh edges, and 854,650 mesh triangles. The AIM grid dimensions are $1000 \times 1000 \times 4$.

The SRR array is excited by a 1 THz plane wave, which has its electric field polarized along the $y$ axis and propagates along the $-z$ axis. The relative tolerance enforced on the residual norm is $10^{-4}$, and the GMRES algorithm takes 43 iterations to converge.
\begin{figure}[!t]
	\centering
	\subfloat{\includegraphics[width=3.45in]{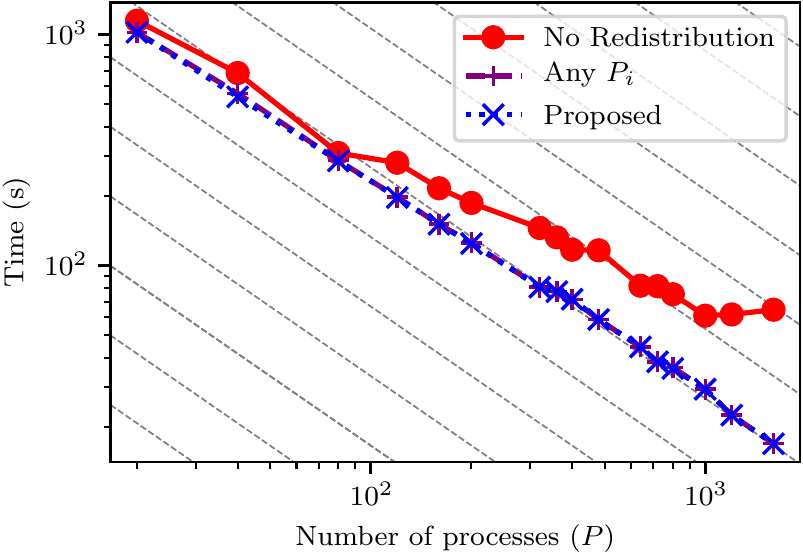}}%
	\\
	\subfloat{\includegraphics[width=3.45in]{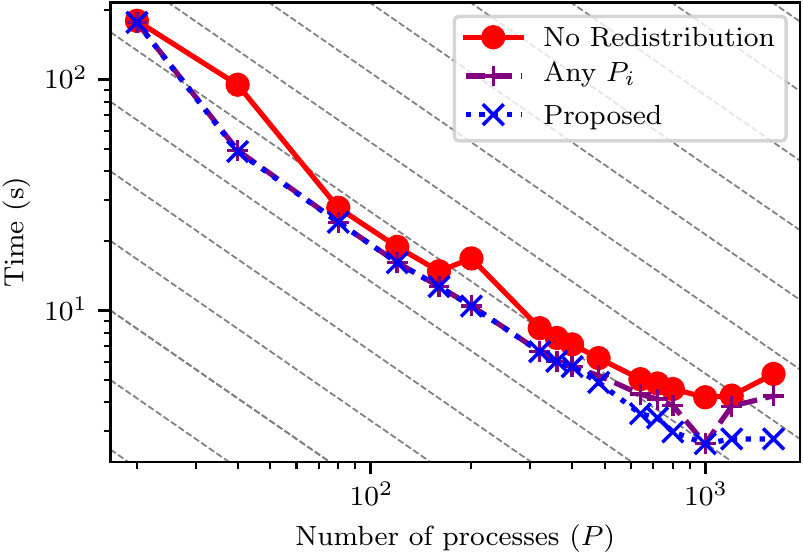}}%
	\caption{Execution times for simulating the SRR array in~\secref{sec:srr} using the proposed solver, with comparisons to the ``No Redistribution'' and ``Any $P_i$'' methods. Gray dashed lines are ideal scalability slopes. Top panel: matrix generation. Bottom panel: total time spent computing matrix-vector products.}
	\label{fig:scala-srr}
\end{figure}

The results of the strong scalability assessment are plotted in~\figref{fig:scala-srr}. The proposed and ``No Redistribution'' methods are compared again, along with the ``Any $P_i$'' method, which is identical to the proposed method but uses the original Part\_Schedule algorithm. The superiority of the proposed method is established. In this example, a smaller fraction of objects are assigned to multiple processes, so the difference between the methods is less pronounced. Nevertheless, the proposed method is up to $3.8\times$ faster at generating the system matrix than the ``No Redistribution'' method. Moreover, when large numbers of processes are used the importance of using approximate squares for $P_i$ is still observed in the matrix-vector product times.

\subsection{Fan-Out Interposer}\label{sec:fan-out}
In this section, the strong scalability of the proposed solver is tested on a structure that contains numerous conductive objects of different sizes and is challenging to load balance. This example structure is inspired by silicon interposers used for 3D integration (courtesy of Dr. Rubaiyat Islam, AMD).

\begin{figure}[!t]
	\centering
	\includegraphics[width=3.49in]{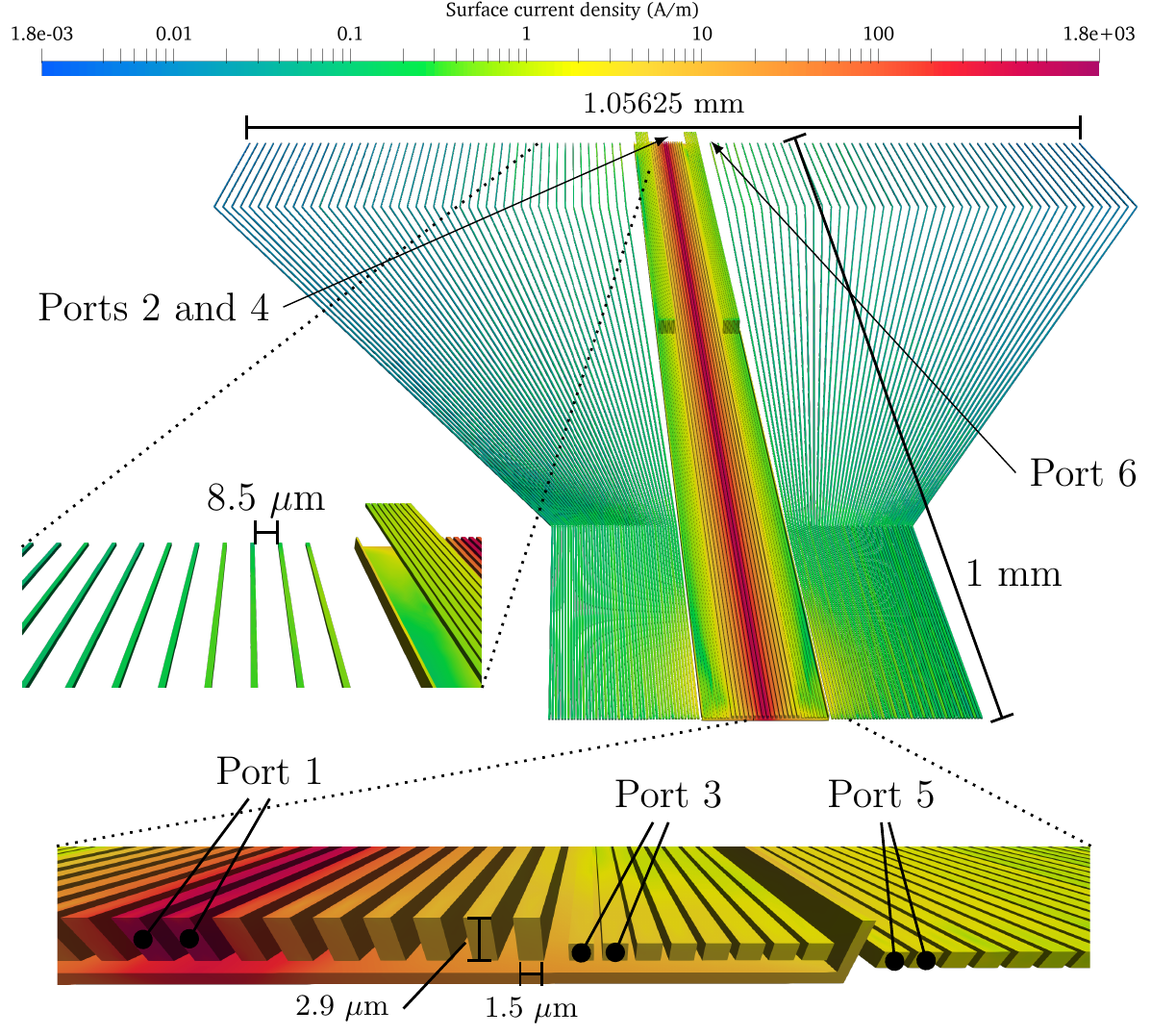}%
	\caption{Geometry of the fan-out interposer with electric current density due to a 10 GHz excitation at port 1.}
	\label{fig:fanout-geom}
\end{figure}

A rendering of the fan-out interposer is presented in~\figref{fig:fanout-geom}. On one end, the spacing between wires ranges from $0.5-1.25~\mu$m, while on the other end wire spacing is an order of magnitude larger, with a maximum spacing of $8.5~\mu$m. In addition, a large ground cage is placed under the central straight-wire section. All objects are made of copper with $\sigma=5.8\times 10^7$ S/m, and the entire structure is embedded in a multilayered medium with properties given in the second part of~\tblref{tbl:laymed}.

The geometric mesh of the fan-out interposer comprises 449,610 edges, 299,740 triangles, and 129 distinct objects. The average mesh element length is $3.2~\mu$m. In addition, the AIM grid is $400 \times 400 \times 8$, with the shortest dimension oriented along the $z$ axis. The interposer is excited at 10 GHz with a total of six lumped ports. Finally, the GMRES algorithm~\cite{Saad1986} is set to terminate after the relative tolerance of the residual norm is below $10^{-4}$. The solver requires on average 76 GMRES iterations per excitation, which is not dependent on $P$.

The large ground cage comprises approximately 53\% of the total estimated internal workload. The remaining 128 objects are smaller lines and their estimated workloads range from 0.3\% to 0.43\% of the total estimated internal workload.

\begin{figure}[!t]
	\centering
	\subfloat{\includegraphics[width=3.45in]{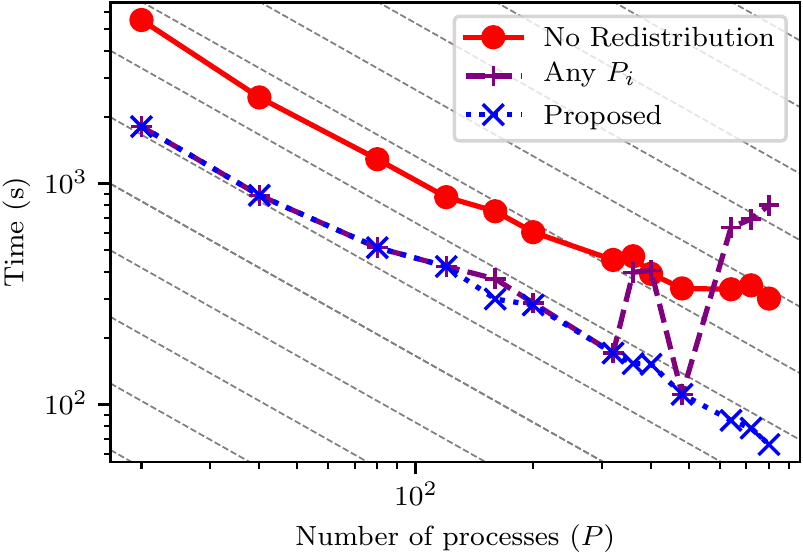}}%
	\\
	\subfloat{\includegraphics[width=3.45in]{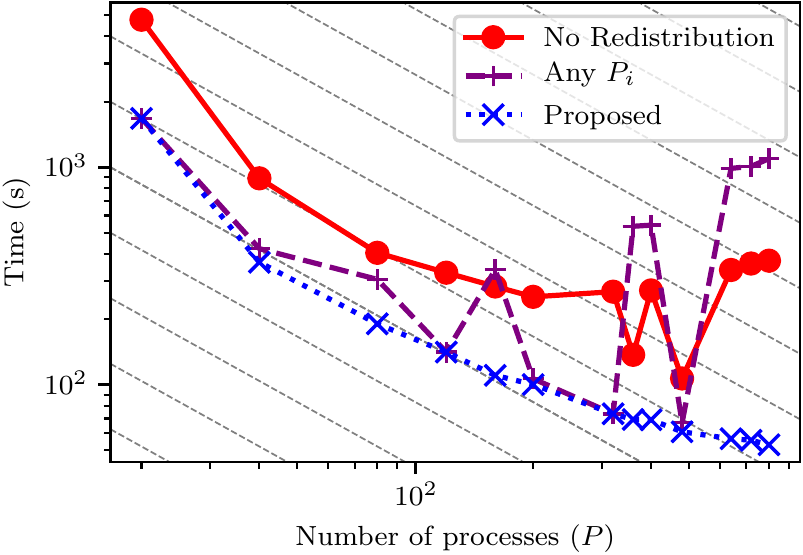}}%
	\caption{Execution times for the fan-out interposer in~\secref{sec:fan-out} using the proposed solver, with comparisons to the ``No Redistribution'' and ``Any $P_i$'' methods. Gray dashed lines are ideal scalability slopes. Top panel: matrix generation. Bottom panel: total time spent computing matrix-vector products.}
	\label{fig:scala-interposer}
\end{figure}

We compare wall times resulting from a strong scalability experiment for the three different methods from~\secref{sec:para-int} in~\figref{fig:scala-interposer}. The plots in~\figref{fig:scala-interposer} clearly show that the ``No Redistribution'' method leads to poor workload balancing in the internal problem. Specifically, the proposed method is as much as $4.6\times$ faster at generating the system matrix and $7\times$ faster at performing the matrix-vector multiplications. Furthermore, the effect of using an arbitrary number of processes on the parallel efficiency of ScaLAPACK is evident in~\figref{fig:scala-interposer}. The performance of the ``Any $P_i$'' method heavily relies on whether the Part\_Schedule algorithm arrives at an approximate square $P_i$ for the large ground cage. These results demonstrate that the parallelization of the internal problem is non-trivial, and naive implementations lead to far-from-optimal performance. In contrast, the proposed scheme provides near-optimal performance while accurately modeling lossy objects.

\section{Conclusion}
We proposed a parallelization strategy for the boundary element method (BEM) applied to Maxwell's equations that, unlike existing parallel solvers, accurately models conductive objects at low frequencies. The BEM formulation adopted in this paper is composed, in part, of a large single computational workload that models electromagnetic coupling between conductive objects and resembles the workload of BEM formulations that use approximate models for lossy conductors. However, incorporating accurate models for conductive objects into the BEM formulation introduces a new set of computational workloads that are coupled to the large existing workload. The proposed method balances these two different workloads in a compatible and efficient manner, which keeps communication costs at a minimum level. In addition, we demonstrate that the challenging problem of balancing the new set of computational workloads can be translated into a parallel task scheduling problem, which is expediently solved by algorithms in scheduling theory.

Three structures with millions of unknowns were used to test the proposed parallelization strategy through weak and strong scalability experiments. The parallel efficiency of the solver remained high when up to 1,000 processor cores were used. In addition, the proposed strategy led to a solver that was up to $7\times$ faster than solvers based on other parallelization methods. The proposed method can be used in conjunction with other acceleration techniques employed in the BEM, e.g., the multilevel fast-multipole algorithm~\cite{Song1995}, and it can be adapted to other formulations of the internal problem, such as, the generalized impedance boundary condition formulation~\cite{Qian2007}.

%

\section*{Acknowledgment}
This work was supported by Advanced Micro Devices, by the Natural Sciences and Engineering Research Council of Canada and by Compute Canada. Computations were performed on the Niagara supercomputer at the SciNet HPC Consortium. SciNet is funded by: the Canada Foundation for Innovation; the Government of Ontario; Ontario Research Fund - Research Excellence; and the University of Toronto.
\ifCLASSOPTIONcaptionsoff
  \newpage
\fi



\bibliographystyle{IEEEtran}
\bibliography{IEEEabrv,ref.bib}


%



%

\begin{IEEEbiography}{Damian Marek}
Biography text here.
\end{IEEEbiography}
\begin{IEEEbiography}{Shashwat Sharma}
	Biography text here.
\end{IEEEbiography}
\begin{IEEEbiography}{Piero Triverio}
	Biography text here.
\end{IEEEbiography}






\end{document}